\documentclass[aps,prmaterials,twocolumn,a4,nofootinbib,showkeys]{revtex4-2}

\usepackage{float}
\usepackage[T1]{fontenc}
\usepackage{graphicx}

\newcommand{\ads}{{\mbox{\scriptsize ads}}}
\newcommand{\surface}{{\mbox{\scriptsize surface}}}
\newcommand{\surf}{{\mbox{\scriptsize surf}}}
\newcommand{\deform}{{\mbox{\scriptsize def}}}
\newcommand{\amino}{{\mbox{\scriptsize amino}}}
\newcommand{\molecule}{{\mbox{\scriptsize adsorbant}}}

\newcommand{\cut}{{\mbox{\scriptsize cut}}}
\newcommand{\fullcell}{{\mbox{\scriptsize full}}}
\makeatletter

\begin{document}

\title{Towards an understanding of magnesium in a biological environment: \\ A density functional theory study}

\author{Miranda Naurin, Sally Aldhaim, Moltas Elliver, Ludwig Hagby, J. Didrik Nilsson, and 
Elsebeth Schr{\"o}der}\email{Corresponding author; schroder@chalmers.se (E. Schr\"oder)}%
\affiliation{Microtechnology and Nanoscience, MC2,
Chalmers University of Technology, SE-41296 G{\"o}teborg, Sweden}

\date{March 2, 2026}
\begin{abstract}

Density functional theory is used to investigate the interactions between a layer of magnesium hydroxide, Mg(OH)$_2$, the magnesium (Mg) surface Mg(0001),
and the three amino acids glycine, proline and glutamine.
The aim is to improve the understanding of Mg behavior in biologically relevant environments, 
such as the ones that  biodegradable implants experience in the body. 
For a simple model of such conditions, adsorption of amino acids are studied. 
With the layer of Mg(OH)$_2$ as a model of either slightly corroded Mg, or intentionally coated Mg, 
the interfacial interaction between a layer of Mg(OH)$_2$ and Mg(0001) is first examined in the absence of the molecules.
Then follows analyses that include amino acids on top of the Mg(OH)$_2$ layer.
We find that the Mg(OH)$_2$/Mg(0001) interaction is weak and that the layer of Mg(OH)$_2$ can readily slide across the Mg surface. 
The presence of amino acids is found to have a limited influence on the adsorption of Mg(OH)$_2$ on Mg(0001), decreasing the
binding by at most 3\%, while more layers of Mg(OH)$_2$ strengthen 
the Mg(OH)$_2$/Mg(0001) binding by 13\%. 
This is still less than the binding of Mg(OH)$_2$ layers within its native bulk structure, and our findings 
indicate that only a small number of hydroxide layers are required before it is energetically 
more favorable for Mg(OH)$_2$ to create bulk than to stay on Mg(0001) as single layers. 
This provides insight into early-stage surface processes relevant for magnesium-based implant materials.

\end{abstract}

\maketitle

\hyphenation{over-estimated mole-cules}


\section{Introduction}

Popular implant materials such as steel and titanium alloys  can provide stable fixation with high mechanical strength \cite{Hung_Kwok_Yip_Wong_Leung_2025}. 
However, for bone healing high strength in the implant is not necessarily an advantage, 
for example, titanium alloys can cause stress shielding to the bone, 
removing most of the stress that the bone would otherwise be subjected to. 
This could cause the bone to evolve more slowly as it needs to be exposed to some stress to strengthen. 
Moreover, if the implant is intended as a temporary support, the implant must be removed in an additional surgery
because these materials are not biodegradable \cite{Tsakiris_Tardei_Clicinschi_2021}.  
This comes with a number of risks for the patient and an additional cost for society. 

Magnesium (Mg) is a candidate for biodegradable implant material. 
It is the fourth most abundant metal in the human body, most of which can be found in the skeletal systems \cite{Kim_See_Li_Zhu_2020}.
Previous studies have demonstrated that Mg implants have good biocompatibility and do not harm bone generation \cite{Kraus_Fischerauer_Hänzi_Uggowitzer_Löffler_Weinberg_2012}, and
in vivo testing has shown that Mg ions may induce bone formation \cite{Witte_Kaese_Haferkamp_Switzer_Meyer-Lindenberg_Wirth_Windhagen_2005}. 
Mg has a density  
similar to that of bone 
\cite{Tsakiris_Tardei_Clicinschi_2021}, while
the density of titanium alloys is significantly larger 
\cite{luthringer2014magnesium}. 
Further, the mechanical properties of Mg are close to those of bone, e.g., 
in vitro studies have shown the elastic modulus of Mg and bone to be similar \cite{Agarwal_Curtin_Duffy_Jaiswal_2016}.

The use of Mg as a biodegradable implant was first introduced in 1906 \cite{Witte_Kaese_Haferkamp_Switzer_Meyer-Lindenberg_Wirth_Windhagen_2005},
but use was abandoned due to the high corrosion rate that resulted in excessive production of bubbles of hydrogen gas (H$_2$).
On the one hand, the released H$_2$ gas in vivo is believed to affect the integration of the implant with the surrounding bone tissue \cite{Kraus_Fischerauer_Hänzi_Uggowitzer_Löffler_Weinberg_2012}, and the gas bubbles can occupy space and potentially induce local mechanical pressure. 
These bubbles can lead to shifts in acidity of the surrounding environment, 
with risk of
negatively affecting the early stages of bone healing, particularly in the outer bone tissue. 
On the other hand, a recent in vivo study conducted on rats \cite{amara25} indicated that such effects might not be critically detrimental. 
Despite elevated degradation rates and substantial hydrogen evolution, complete bone regeneration was still achieved. 

Still, the stability and mechanical properties of the implant must be preserved for 
a certain period of time on the scale of a year.
With this in mind, controlling the corrosion rate of magnesium is of high priority and a necessary step towards making magnesium-based biodegradable implants. 
Approaches include the alloying of Mg or coating the Mg surface with a less reactive material \cite{Kim_See_Li_Zhu_2020,Agarwal_Curtin_Duffy_Jaiswal_2016}.

In the present work we use density functional theory (DFT) calculations to study the adsorption of a layer of magnesium hydroxide, Mg(OH)$_2$, 
on top of the Mg(0001) surface, modeled as an intentional coating or as a result of the first stages of Mg oxidation. 
We determine the energy needed to peel off the Mg(OH)$_2$ layer directly, and the energy needed to slide the layer across the Mg surface, which would eventually
lead to the layer sliding off the Mg surface.
Further, we calculate the adsorption energies and geometries of the body-relevant amino acids glycine (Gly), proline (Pro) and glutamine (Gln) on this coating, 
and study how they influence the binding of Mg(OH)$_2$ to Mg(0001). 

The text is organized as follows. 
Section II introduces the methods used and materials studied, Section III contains our
results including discussions, and conclusions and a summary are given in Section IV.
Finally, in Appendix \ref{param} we discuss convergence of central parameters in the calculations.


\section{Methods and Materials}
\subsection{Density functional theory calculations} 
DFT is used to calculate the structures and energies of the adsorbed coating and amino acid molecules.
We employ the plane wave code \texttt{pw.x} of the Quantum ESPRESSO suite \cite{QE,espresso,QE_2}, with 
exchange and correlation approximation given by the vdW-DF-cx 
method \cite{dion04p246401,schroder17chapter,berland14p035412,berland14jcp}. 
Since some of the adsorption is expected to be weak, it is important to use an approximation that self-consistently 
includes dispersion, or van der Waals (vdW) interactions, such as vdW-DF-cx. The vdW-DF-cx has current conservation as
guidance and has been shown to work 
excellent in adsorption of organic molecules \cite{berland15p066501,berland14jcp}.

We use PAW-based \cite{PAW} pseudopotentials (PP) that were all created by the \texttt{ld1.x} code \cite{dalcorso14} of Quantum ESPRESSO. 
For Mg we use the two-valence-electron PP Mg-pbesol-paw-3d.UPF, created and tested by one of us in Ref.~\cite{schroderPP}, while all
other PPs are from the PSLibrary package \cite{pslibrary}. 

As described in Appendix \ref{param}, we tested and found that a dense (more sparse) Monkhorst-Pack \cite{monkhorst76p5188} k-point 
mesh with $18\times 18\times1$ ($10\times10\times1$) points
in the $1\times1$ surface cell is sufficient for an accuracy in energy difference of 0.02 mRy (0.2 mRy) per unit cell. 
For all calculations of Mg(OH)$_2$ on top Mg(0001) we use the dense mesh.
For calculations  of amino acid adsorption we use the larger $5\times5$
surface cell.
The structures and thus the adsorption energies of the adsorbed molecules are inherently 
less accurate (due to multiple local minima in the energy landscape, of which we can never catch all).
It therefore suffices to use the more sparse mesh for these calculations, 
corresponding to $2\times2\times1$ k-point in the $5\times 5$ surface cell. 

Similar tests for the cut-off energy values of the plane-wave kinetic energy $E_\cut$ and electron densities $E_\cut^\rho$  yield 
$E_\cut =50$ Ry and $E_\cut^\rho =400$ Ry for sufficient accuracy in the present project.
For the graphene and graphite calculations we use corresponding parameters.
For more details on our convergence tests  we refer to Appendix \ref{param}.

Unless otherwise stated, the atomic positions in the system are optimized such that the Hellmann-Feynman forces, obtained 
from the DFT-generated electron charge density, are minimized. 
This does not apply to the lower three Mg layers in the Mg(0001) surface, as
explained in the next subsection.

Finally, for testing a number of realistic starting configurations of the amino acid positions on the surface, we employed
the molecular editor Avogadro \cite{avogadro}
using (classical) molecular mechanics  to obtain a series of fast, guided, 
initial guesses for the adsorption geometries, to use as input 
to DFT calculations with minimization of the Hellmann-Feynman forces to
further optimize the atomic positions.

\subsection{Magnesium surface and Mg(OH)$_2$}

Mg(OH)$_2$ is a known corrosion product of Mg surfaces in contact with a biological 
environment \cite{petrovic20,willumeit11,tomozawa11}.
It is therefore expected to naturally be present on or near Mg-based implants but it can also be 
 part of an intentionally created coating \cite{tshizaki13,zhang14}.
At ambient temperature and pressure Mg bulk condenses in a hexagonal closed packed (HCP) structure. 
Mg(OH)$_2$ is a layered material which also has the HCP structure, 
with a lateral lattice constant only slightly smaller than that of bulk Mg.
This means that we can place a layer of Mg(OH)$_2$ on top Mg(0001) with the same periodicity, imposing only a minor strain on  Mg(OH)$_2$.

When only Mg(OH)$_2$ is adsorbed on Mg(0001) we thus use a 1$\times$1 surface unit cell of Mg(0001). 
However, when amino acid molecules are adsorbed on top, we use the laterally larger $5\times 5$
surface cell, to avoid spurious interactions of the molecules across the periodic boundaries. 
For the largest molecule studied here (Gln) the smallest distance between atoms on two periodic copies of the molecule is then 9~{\AA}.
The lattice constants of Mg bulk, with the PP used here, were found in Ref.\ \onlinecite{schroderPP}, 
and thus our 1$\times$1 Mg(0001)-surface is created with lateral lattice constant 
$a=3.192$ {\AA}. 

We model Mg(0001) using five Mg layers, of which the lowest three are kept in the positions they had in a fully relaxed 23-layer Mg slab \cite{schroderPP,bolin26}. 
All other atoms, including atoms of the overlayer and molecules, are allowed to relax, unless specified.
For the direction perpendicular to the surface 
we use a unit cell height of 36 {\AA}, 
leaving at least 14 {\AA} of vacuum even when amino acids are adsorbed.
In the vacuum region
we impose a dipole correction to account for the top and bottom of the surface slab being different \cite{bengtsson99p12301}.

For bulk Mg(OH)$_2$ we find the lateral lattice constant to be 3.130 {\AA} 
(experimental value  $a= 3.15$ {\AA} \cite{catti95}),
which means that we
must expand Mg(OH)$_2$ by 2\% in order to fit a layer of Mg(OH)$_2$ on top of Mg(0001). 

\subsection{Adsorption energies}
The strength of the binding of Mg(OH)$_2$ to Mg(0001) and of the amino acids on Mg(OH)$_2$ on top of  Mg(0001)
is quantified by the adsorption energy $E_\ads$, defined as
\begin{equation}
E_\ads = E_{\fullcell}-E_\surface-E_\molecule \, .
\label{eq:ads}
\end{equation}
Here $E_{\fullcell}$ is the calculated total energy for the full system, 
and $E_{\surface}$ and $E_{\molecule}$ are those of the surface and the adsorbant.
Depending on the adsorption energy of interest, the surface is either Mg(0001) or Mg(0001) with Mg(OH)$_2$,
and the adsorbant is either a  Mg(OH)$_2$ layer (with or without an amino acid) or an amino acid molecule. 
For both  $E_\surface$  and $E_\molecule$ we use the relaxed structures, allowing also Mg(OH)$_2$ 
to release the lateral strain that is imposed at adsorption, to fit to the Mg(0001) lattice constant.
With the definition in (\ref{eq:ads}) binding is obtained for negative values of $E_\ads$.

$E_\ads$ is the energy required to lift the molecule or Mg(OH)$_2$ layer directly from the surface. 
However, if the Mg(OH)$_2$ layer can slide on Mg(0001) it might eventually slide off, possibly at a lower energy cost than directly lifting it off.
We therefore also calculate the difference in $E_\ads$ for Mg(OH)$_2$ as the layer is moved across Mg(0001), 
creating a potential energy surface (PES) plot.
The largest difference in $E_\ads$ for any two positions of Mg(OH)$_2$ on Mg(0001), $\Delta E_\ads$, is an upper limit on the energy 
needed for sliding the layer, and it can be taken as an estimate of the corrugation of Mg(0001) for Mg(OH)$_2$ \cite{akesson12p174702}. 
For the PES plot we need $E_\ads$ calculated for Mg(OH)$_2$ also when it is not in a (local) optimal position. 
We therefore keep the lateral 
positions of the atoms in Mg(OH)$_2$ fixed, to avoid the layer from sliding back to the (local) minimum during the optimization 
of the atomic positions. This will be further discussed in the results section.

\subsection{Amino acids}
We study the adsorption of three different amino acids on the Mg(OH)$_2$-covered surface. 
The molecules Gly and Pro are two common components of type-1 collagen. 
Type 1-collagen is one of the most common biomolecules in the human body, and it plays a crucial role in the formation of bodily tissue due to its highly stable triple-helix structure \cite{Naomi_Ridzuan_Bahari_2021}. 
Each string consists of a repeating sequence of the amino acids Gly-X-Y, where  X and Y are various amino acids, most commonly Pro and hydroxyproline. 

Collagen constitutes the primary component of the extracellular matrix, which is essential for bone tissue regeneration \cite{Nie_Sun_Wang_Yang_2020}. 
The extracellular matrix is the substance located between cells and it contributes to numerous biological functions within the body. 
Of particular relevance to this project, studies have demonstrated that Mg ions can bind to collagen, thereby promoting osteoblast differentiation through specific biological signaling pathways. 
Osteoblasts are bone-forming cells responsible for stimulating the production of new bone tissue, and are therefore critical in the regeneration process following a fracture \cite{Theocharis_Skandalis_Gialeli_Karamanos_2016}. 
Collagen also contributes to the structural integrity of both cartilage and bone, making its binding interactions with potential implants a subject of significant interest. 

Gln is the third
amino acid used in this project. 
In cells of skeletal muscle more than 50\% of  
unbound amino acids are Gln \cite{Hall_Heel_McCauley_1996} and 
the skeletal muscle is the main tissue for Gln release, synthesis and storage \cite{Cruzat_2019}. 
Hence, Gln is of high relevance to this project.


\section{Results and discussion}

We first focus on the adsorption of a layer of Mg(OH)$_2$ on Mg(0001) and the search for the optimal adsorption geometry.
Thereafter, amino acid adsorption is discussed, along with the effect of this on the Mg(OH)$_2$/Mg(0001) binding. 

\begin{figure}[tb]
  \centering
   \includegraphics[width=0.32\linewidth]{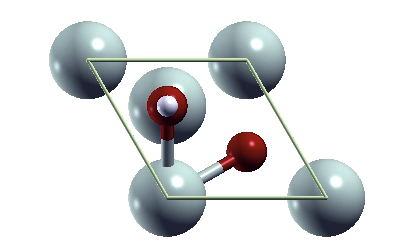}
  \hfill
    \includegraphics[width=0.32\linewidth]{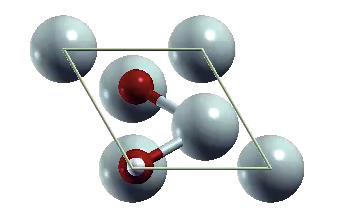}
  \hfill
    \includegraphics[width=0.32\linewidth]{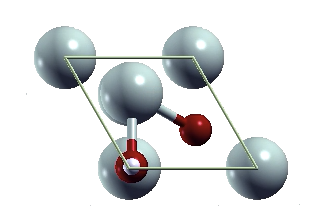}\\
      (a) \hspace{0.25\linewidth} (b) \hspace{0.25\linewidth} (c) 
\caption{\label{fig:mgoh2_on_mg}Mg(OH)$_2$ on top of Mg(0001) in a $1\times1$ surface unit cell. 
  (a) Mg(OH)$_2$ with the lower H atom above an FCC hollow site on Mg(0001) and Mg of Mg(OH)$_2$ above an HCP top site;
  (b) the optimal position for Mg(OH)$_2$, with H above an HCP hollow site and Mg above an FCC hollow site;  
  (c) the lower H atom above an FCC hollow site  and Mg of Mg(OH)$_2$ above an HCP hollow site; this
  position cannot be obtained by simply translating Mg(OH)$_2$ in (a) without rotations, the adsorption energy was separately calculated. }
\end{figure}

\subsection{Mg(OH)$_2$ on Mg(0001)}
For a layer of Mg(OH)$_2$ adsorbed on Mg(0001), 
one lateral unit cell of  Mg(OH)$_2$ fits on one lateral unit cell of Mg(0001) if stretched by 2\% 
in the lateral directions.  
However, we do not a priori know the optimal relative position of the Mg(OH)$_2$ layer,
because local atoms of the layer and the surface will interact.

The Mg(OH)$_2$ layer has a H atom pointing towards the surface. 
Previous studies have shown that for a single atom of H adsorbed on Mg(0001) the energetically optimal position is in the FCC hollow site \cite{jacobson02,jiang10}, as also tested and found in the present study. 
One might therefore expect that the lateral position of the overlayer is such that the lower H of Mg(OH)$_2$ is above the FCC hollow site in Mg(0001), with two examples illustrated in Figure \ref{fig:mgoh2_on_mg}(a) and (c).
However, H as part of Mg(OH)$_2$ carries a different charge than a single H atom, with the major part of the 
electron density moving from H to the O atom in the O-H binding.
Other relative positions may therefore be 
more favorable. 

\begin{figure}[tb] 
\begin{center}
\includegraphics[width=\columnwidth]{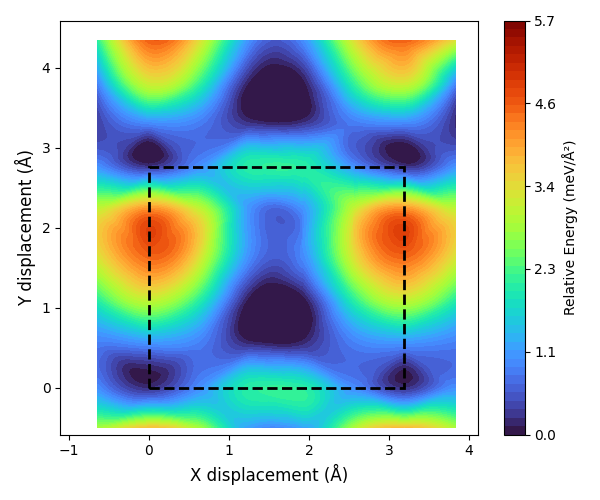}
\end{center}
\caption{\label{fig:heatmap} Potential energy surface plot for a layer of Mg(OH)$_2$ translated across Mg(0001). The dashed black box illustrates the periodicity of the displacement. 
The energy is given relative to the optimal position. 
The optimal position, found in the dark-blue region 
close to the center of the dashed box, is the position shown in Figure \ref{fig:mgoh2_on_mg}(b).}
\end{figure}

We carry out a study 
to determine the adsorption energy $E_\ads$ in a grid of relative positions across the surface, as described below.
The Mg(OH)$_2$ is translated in the lateral directions (without any rotation). 
By this we obtain (an estimate of) the PES of Mg(OH)$_2$ on Mg(0001), shown in Figure \ref{fig:heatmap}. 
From the PES we find that instead of H taking up the FCC hollow site, Fig.~\ref{fig:mgoh2_on_mg}(a), 
the Mg atom of Mg(OH)$_2$ does. 
The optimal geometry is shown in Figure \ref{fig:mgoh2_on_mg}(b), with the lowest H atom positioned above a
second-layer Mg atom in Mg(0001), in a HCP hollow site.

The PES in Figure \ref{fig:heatmap} is obtained by translating Mg(OH)$_2$ across the surface.
The layer is moved in equal increments in each lateral direction,
resulting in 25 calculations  within one unit cell of the Mg surface.
In each lateral point on the surface, $\mathbf{r}_\parallel$, the two top layers of Mg atoms in Mg(0001) 
are free to move, while in Mg(OH)$_2$ the atoms are only free to move in the direction perpendicular to the surface. 
The adsorption energies thus found are estimates of the correct adsorption energies, because the atoms of 
Mg(OH)$_2$ in the PES are restricted in their degrees of freedom.
Figure \ref{fig:heatmap} shows the results of these calculations, $\Delta E_\ads(\mathbf{r}_\parallel)$, with the energy scale relative to the adsorption energy in the optimal position.   

Figure \ref{fig:heatmap} shows that the
adsorption energy differences are relatively small,
with the largest energy difference 5.7 meV/\AA$^2$.
This indicates that the layer of Mg(OH)$_2$ easily can slide off of the Mg surface, 
almost as easily as a layer of graphene can slide off graphite, as seen in Table \ref{tab:graphite}.
It also suggests a stability against rupture due to strain for large patches on Mg(0001): The cost
of having parts of the patch sitting on less favourable positions if the strain is partly released
is small.

Based on the estimates of adsorption energies in Fig.~\ref{fig:heatmap}, we calculate more accurate adsorption energies in the global minimum, corresponding to Fig.~\ref{fig:mgoh2_on_mg}(b), and in the two structures corresponding to Fig.~\ref{fig:mgoh2_on_mg}(a) and (c), 
now with the atoms of Mg(OH)$_2$ allowed to relax in all directions.
We find the optimal adsorption energy, as defined by Eq.~(\ref{eq:ads}), to be $E_\ads=-17.9$ meV/{\AA}$^2$.
This is a relatively small value compared, for example, with the energy cost of lifting off a layer of graphene from graphite, 
which we calculated to be $-25.3$ meV/{\AA}$^2$, Table \ref{tab:graphite}. 
For the two other (metastable) structures 
we find $E_\ads= -17.1$ meV/{\AA}$^2$ (Fig.~\ref{fig:mgoh2_on_mg}(a)) and $-17.6$ meV/{\AA}$^2$ (Fig.~\ref{fig:mgoh2_on_mg}(c)).

Above, we studied the adsorption energy relative to a free-floating layer of Mg(OH)$_2$, with atomic positions and
lateral lattice constant optimized. 
If we evaluate instead the energy relative to a layer of Mg(OH)$_2$ inside its bulk, we find 
the adsorption energy between the Mg-surface and Mg(OH)$_2$-layer to be $+314$ meV per surface unit cell, or
$+36.5$ meV/\AA$^2$, as further discussed below. 
The fact that the result is positive indicates that a layer of Mg(OH)$_2$ on Mg(0001) 
is not  created spontaneously from bulk.
However, corrosion reactions can create Mg(OH)$_2$  and that is a more realistic route to
Mg(OH)$_2$ on Mg(0001).
Further, we see that if in the corrosion process several layers of Mg(OH)$_2$ are formed, they bind more strongly to each other than 
one or more  Mg(OH)$_2$ layers on Mg(0001). This is discussed further in Section III.C.

\begin{figure*}[tb]
\begin{center}
\includegraphics[width=1.6\columnwidth]{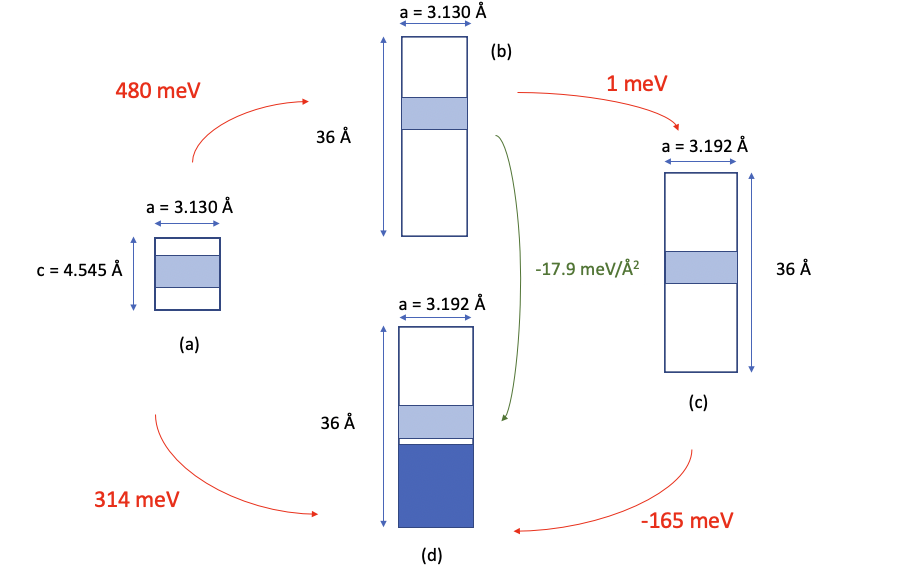}
\end{center}
\caption{\label{fig:Mg-cykel}
Illustration of the energies that go into creating a Mg(OH)$_2$ layer on Mg(0001) either from bulk Mg(OH)$_2$, (a) to (d), 
or from an isolated layer of Mg(OH)$_2$, (b) to (d), in an imagined separation of the process.
Light blue illustrates the Mg(OH)$_2$-layer and dark blue the Mg-surface. 
Values at the red and green arrows indicate the energy required to go from one step to the other. All energies in meV per surface unit cell, except the (b) to (d) energy.
}
\end{figure*}

Figure \ref{fig:Mg-cykel} illustrates energies involved in an imagined step-by-step process
for creating a layer of Mg(OH)$_2$ from bulk. First, the layers would need to be separated, 
this is calculated as sketched in Figure \ref{fig:Mg-cykel}, going from (a) to (b). 
Subsequently, to fit with the periodicity of Mg(0001) the lateral lattice constant $a$ of Mg(OH)$_2$ must be increased,
Fig.~\ref{fig:Mg-cykel}(b) to (c).
Finally, the layer is placed on Mg(0001), with relaxations of the atomic positions, Fig.~\ref{fig:Mg-cykel}(c) to (d). 
Each step except the last requires energy, as illustrated by the energies per lateral unit cell
provided in Figure \ref{fig:Mg-cykel}.
Our definition of $E_\ads$ corresponds to going from (b) to (d) in Figure \ref{fig:Mg-cykel}, while creating an overlayer from
bulk Mg(OH)$_2$ requires the full cycle from (a) to (d).
    
As shown by the energies in Figure \ref{fig:Mg-cykel},
separating the layers of Mg(OH)$_2$ costs the most energy. 
Some energy is gained when Mg(OH)$_2$ layers are placed on the Mg-surface but not sufficient for the whole process to be energy-efficient. 

We also observe that changing the lateral lattice constant for Mg(OH)$_2$ from the bulk and free-floating value $a= 3.130$ {\AA}
to the 3.192 {\AA} of Mg(0001) barely has an impact on the energy, Figure \ref{fig:Mg-cykel} (b) to (c). 
This means that stretching the Mg(OH)$_2$-layer from its preferable lateral size does not have a significant impact on the system,
thus confirming that it is reasonable to simplify the calculations by letting one unit cell of Mg(OH)$_2$ correspond to one surface unit cell of Mg(0001).

\begin{table}[tb]
    \caption{Calculated energy differences between graphene on graphite and Mg(OH)$_2$ on magnesium. 
    $E_\ads$ is the energy required to lift off the top layer and $\Delta E_\ads$ is the maximum 
    energy required for the top layer to slide. Values in meV/{\AA}$^2$.}
    \begin{center}
    \begin{tabular}{l|cc}
        System & $E_\ads$ & $\Delta E_\ads$\\
        \hline
        Graphene on graphite & -25.3 & 3.7 \\
        Mg(OH)$_2$ on magnesium & -17.9 & 5.7\\ 
    \end{tabular}
    \label{tab:graphite}
    \end{center}
\end{table}

\subsection{Amino acid adsorption}

\begin{table}[tb]
    \centering
    \caption{Adsorption energy $E_\ads$ between the amino acid and the surface (Mg(OH)$_2$ on Mg), 
    as well as the resulting deformation energy $E_\deform^\amino$ and $E_\deform^\surf$ of each component. 
    `(weak)' and  `(strong)' denote the weakly and 
    strongly chemisorbed molecules. Values in eV per molecule.}
    \small
    \begin{tabular}{l|ccc}
        Calculations & $E_{\ads}$ & $E_\deform^\amino$  & $E_\deform^\surf$  \\
        \hline
        Proline(weak) & -0.445     & 0.010   & 0.259    \\
        Proline(strong) & 
          -0.741  &  3.001  &  0.919   \\
        Glycine(weak) & 
        -0.339 & 0.086  & 0.287   \\ 
        Glycine(strong) & 
        -0.892 & 3.054  & 1.281   \\ 
        Glutamine & 
        -0.681   &  0.091  & 0.381   \\ 
\end{tabular}
    \label{tab:deformation}
\end{table}

We study the adsorption of single amino acid molecules on 
top of the Mg(OH)$_2$ layer, at low coverage (one molecule per $5\times5$ Mg(0001) unit cell). At
adsorption both the molecule and the layer will adjust their atomic positions.
We study this change in the geometry, in the corresponding energy changes, 
and in the next Section we discuss the possible secondary effect of amino acid adsorption affecting the binding of Mg(OH)$_2$ on Mg(0001).

Table \ref{tab:deformation} shows, by the size of the deformation energies $E_\deform^\amino$ and $E_\deform^\surf$, that there is  deformation both on the amino acids and on the surface when in contact, though mostly (as expected) when the molecule is
strongly chemisorbed to Mg(OH)$_2$.

\begin{figure}[tb]   
   \centering
        \includegraphics[width=0.48\linewidth]{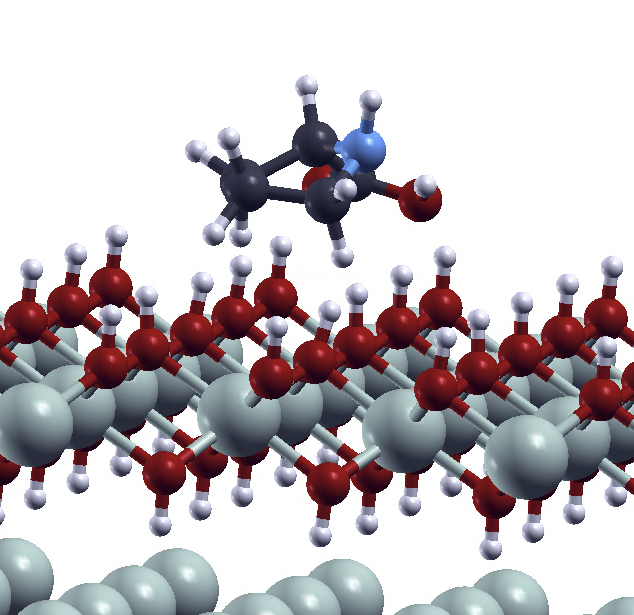}
    \hfill
        \includegraphics[width=0.48\linewidth]{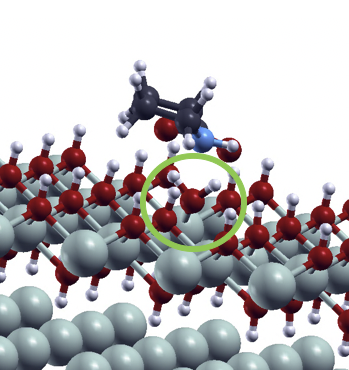}\\
        (a) \hspace{0.40\linewidth} (b)  \\
        \includegraphics[width=0.48\linewidth]{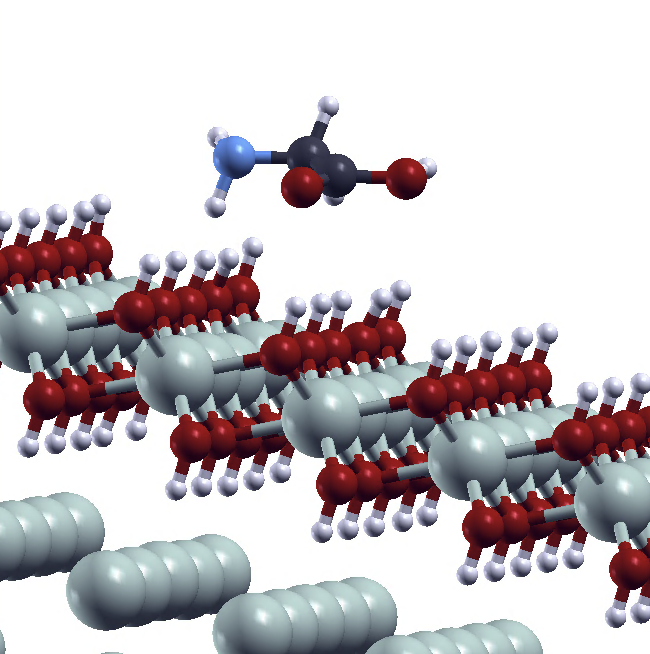}
    \hfill
        \includegraphics[width=0.48\linewidth]{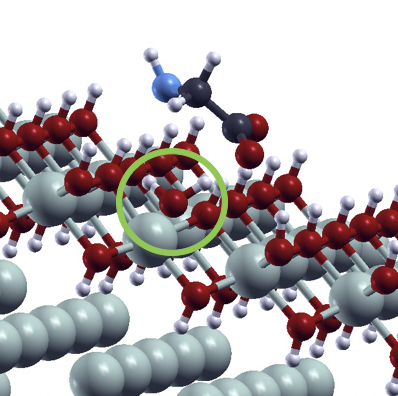} \\
        (c) \hspace{0.40\linewidth} (d)  \\
        \includegraphics[width=0.48\linewidth]{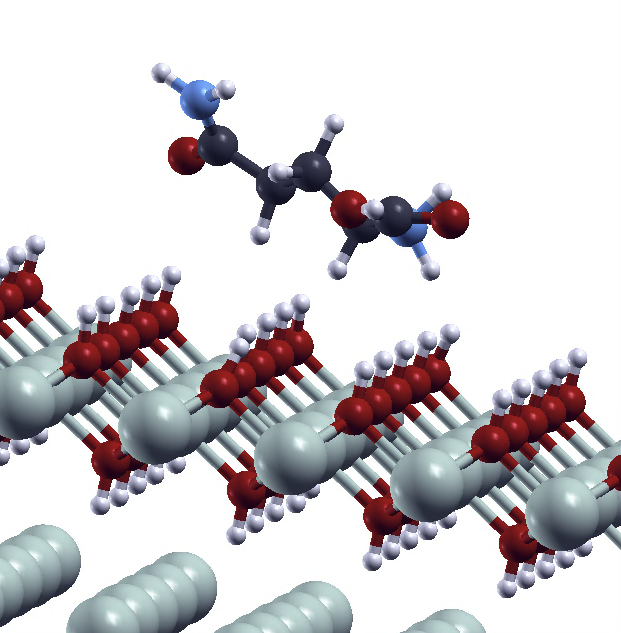}\\
        (e)
    \caption{\label{fig:amino_on_mgoh2} Amino acids interacting with Mg(OH)$_2$ on Mg(0001):
    (a) weakly chemisorbed proline,
    (b) strongly chemisorbed proline, 
    (c) weakly chemisorbed glycine, (d) strongly chemisorbed glycine, (e) glutamine.
    The circles in (b) and (d) indicate the position of the H atom dissociated from the amino acid.}
\end{figure}

The largest impact on both the surface and the molecule arises from the surface interaction with strongly chemisorbed Gly and Pro.
At a closer look at the structures, Fig.~\ref{fig:amino_on_mgoh2}(b) and (d),
we find that in the two molecules an H from an OH group dissociates in the adsorption. 
The weaker adsorption of Gly and Pro, Fig.~\ref{fig:amino_on_mgoh2}(a) and (c), and the adsorption of Gln, 
Fig.~\ref{fig:amino_on_mgoh2}(e), on the other hand, does not give rise to any large deformations
although the deformation energy of the surface is larger than that of the molecule, especially for Pro.
We judge this from the deformation energies in Table \ref{tab:deformation} and the adsorption structures in Fig.~\ref{fig:amino_on_mgoh2}.

Below we discuss each of the amino acids and their adsorption properties.

\textit{Proline.}
The structure of strongly chemisorbed Pro is illustrated in Figure \ref{fig:amino_on_mgoh2}(b), with the circle 
highlighting the H atom dissociated from Pro.
Upon adsorption, the strongly chemisorbed Pro positions itself such that the H atom of one of its OH-groups can 
interact with an OH-group on the Mg(OH)$_2$-layer. 
The H-atom in -OH on Pro detaches, and forms a water-like structure with the OH-group that binds to Mg. 
The O-Mg separation increases from 2.113 {\AA} for O to the three neighboring Mg atoms
in undisturbed Mg(OH)$_2$ 
to 2.199 {\AA} separation from one Mg atom and no bond (distances exceeding 3.5 {\AA}) to the two other Mg atoms.
The dissociation and water formation means that the deformation energies in Table \ref{tab:deformation} also include the dissociation energy of the H-atom from Pro and the change in the -OH group on Mg(OH)$_2$. 
The more weakly chemisorbed Pro does not deform the surface or itself as strongly.
While the deformation energies differ much between the weakly and strongly chemisorbed situation for Pro,
the difference in $E_\ads$ is only
$\approx 300$ meV.

\textit{Glycine.}
Gly also adsorbs both in a weak chemisorption, Fig.~\ref{fig:amino_on_mgoh2}(c), and in a stronger 
chemisorption with dissociation of an H atom from
the -OH group on Gly, Fig.~\ref{fig:amino_on_mgoh2}(d).
Both structures are stable, but the deformation energies involved in the dissociation are, as expected, 
more than an order of magnitude larger, see Table \ref{tab:deformation}.
Again, a water-like structure forms on Mg(OH)$_2$ by the H from Gly attaching to an -OH group on Mg(OH)$_2$,
with O-Mg distance changing to 2.198 {\AA}, similar to the Pro case. In Ref.\ \cite{tang23} the adsorption of zwitterionic glycine
on Mg(OH)$_2$ was studied. Due to the ionic character of the two ends of the molecule the adsorption energy found 
there was larger, at $\approx 3$ eV/molecule.

\textit{Glutamine.}
For Gln, we find a number of similar adsorption structures, illustrated by one of them in Figure \ref{fig:amino_on_mgoh2}(e).
However, despite many different starting positions for the optimization of the adsorption structure, 
including some with the OH group placed close to the surface, none of the calculations
result in dissociation of the H atom from Gln.
Further, Table \ref{tab:deformation} shows that there is merely a small change in the deformation energy
of both the surface and Gln when they interact.
In other words, the adsorption and deformation energies remain the same or similar at the end of each calculation regardless of how the molecule is initially positioned. 

Gly and Pro are smaller  molecules than Gln and face less steric hindrance at adsorption, 
making both weak and strong chemisorption possible. 
This is also seen when Gly adsorbs directly on the metal surface of Mg(0001) \cite{bolin26}, 
although in that study the difference in 
adsorption energy for the two cases is minute.

The dehydration of Mg(OH)$_2$ that Pro and Gly thus contribute to is interesting, because 
in itself (no added amino acids) this process has a very slow reaction rate at body 
temperatures \cite{youn24}, but Pro and Gly adsorption increases the chances.
At the same time, dehydration could result in the formation of a pit in Mg(OH)$_2$, 
exposing the Mg surface to the environment and the start of a corrosion pit in Mg.

\begin{table}[tb]
    \centering
    \caption{Adsorption energy of a $5\times 5$ layer of Mg(OH)$_2$ adsorbed on Mg(0001), with one molecule of the 
    indicated amino acid or a second layer of Mg(OH)$_2$ adsorbed on top of Mg(OH)$_2$. 
    Values in eV per adsorbed amino acid (and $5\times 5$ surface unit cell). }
    \small
    \begin{tabular}{lc}
        Amino acid on Mg(OH)$_2$/Mg(0001) & $E_{\ads}$\\
        \hline
        none & -3.95 \\ 
        Proline(weak) & -3.87\\
        Proline(strong) & -3.86 \\ 
        Glycine(weak) & -3.86\\
        Glycine(strong) & -3.83 \\
        Glutamine & -3.87 \\ 
        Second layer Mg(OH)$_2$ & -4.45 \\ 
    \end{tabular}
    \label{tab:ads.energy_Mg}
\end{table}

\subsection{Effect on the Mg(OH)$_2$/Mg(0001) interaction}
With the amino acids adsorbing this differently on Mg(OH)$_2$ one could expect 
the effect to progress further into the surface, to affect how the Mg(OH)$_2$ layer binds to Mg(0001).
One could also expect that the effect would depend on which amino acid, if any, is adsorbed on top, as well as
the strength of the adsorption.

Table \ref{tab:ads.energy_Mg} reports the difference in adsorption energies of Mg(OH)$_2$ on Mg(0001). 
The results shows that 
i) adsorbing an amino acid on Mg(OH)$_2$ weakens the Mg(OH)$_2$-Mg(0001) binding energy by less than 120 meV per 
$5\times 5$ unit cell (3\%) and thus per molecule, and 
ii) the type of amino acid and strength of adsorption on the Mg(OH)$_2$-layer hardly matters for the adsorption energy 
between Mg(OH)$_2$ and Mg(0001), at less than 40 meV/molecule, or 1\%. 
Thus, despite the change in atomic and electron structure on the top of Mg(OH)$_2$ when Gly or Pro provides 
an H atom in strong chemisorption, 
it does not affect the binding of Mg(OH)$_2$ to Mg(0001) noteworthy.

A second layer of Mg(OH)$_2$ on top of the initial layer has the effect that Mg(OH)$_2$ binds stronger to Mg(0001), but only about 500 meV per $5\times5$ surface unit cell (13\%) compared to a single layer on Mg(0001),
while the Mg(OH)$_2$-Mg(OH)$_2$ binding is much stronger, at 12 eV for the same area 
(i.e., 480 meV per $1\times1$ surface cell).
These results also show that it only requires a few layers of Mg(OH)$_2$ before it is energetically much better 
in bulk rather than adsorbed to Mg. 
While the Mg(OH)$_2$ layer on Mg(0001) may have an entirely different origin than being cut out from bulk Mg(OH)$_2$ (e.g., created on top 
of the Mg surface by corrosion of Mg), 
once more Mg(OH)$_2$ layers are formed one can envision a reverse process: 
multiple layers of Mg(OH)$_2$ lifting off the Mg(0001) surface.

\section{Summary} 
This study provides a detailed analysis of the interface of a layer of Mg(OH)$_2$ with Mg(0001)  and 
the changes at adsorption of amino acids that are relevant in the environment of Mg-based implants.
Via a potential energy surface plot we find the optimal adsorption position of Mg(OH)$_2$ on Mg(0001).
The associated adsorption energy
is $-17.9$ meV/{\AA}, and the corrugation, estimated from the difference in largest and smallest 
adsorption energy of any position, is small with value 5.7 meV/{\AA}.
We thus conclude that Mg(OH)$_2$ is both relatively easy to peel off (at lower adsorption energy than graphene on
graphite) and can easily slide on top of Mg(0001).

Our study further shows that glycine, proline and glutamine interact with the Mg(OH)$_2$ layer to varying extent, 
with glycine and proline adsorbing in both weak and strong chemisorption. 
For glycine and proline the strong chemisorption increase the deformation energies of Mg(OH)$_2$ and the molecule
itself significantly, while the adsorption energies are less affected. 
This increase is attributed to the dissociation of an H atom from the amnio acid OH group, moving to an OH group on Mg(OH)$_2$
to form a water-like structure, though still attached to one Mg atom in Mg(OH)$_2$. 
In contrast, glutamine does not show dissociation of its OH group. 

The presence of amino acids on Mg(OH)$_2$ is found to have little impact on the adsorption energy of Mg(OH)$_2$ on Mg(0001).
Adding another layer of Mg(OH)$_2$ affects the Mg(OH)$_2$-Mg(0001) binding more, but the strength of Mg(OH)$_2$-Mg(0001)
stays at orders of magnitude smaller than the binding energy of layers within bulk of Mg(OH)$_2$-Mg(0001).

Our work shows that while Mg(OH)$_2$ may form due to corrosion of the Mg surface, it is relatively easy to remove and this does not
change significantly with interaction of amino acids or formation of further Mg(OH)$_2$ layers. 
Our study likewise shows 
that Mg(OH)$_2$ does not project the Mg surface sufficient for an efficient coating by itself.

\begin{table}[tb]
    \caption{Calculated energy differences $\Delta E$ between Mg(OH)$_2$ adsorbed on Mg(0001) and the 
    layer translated to 0.1 {\AA} above Mg(0001), without further relaxing the atomic positions. 
    Values are per ($1\times1$) surface unit cell.}
    \begin{center}
    \begin{tabular}{l|c}
        k-points  &$\Delta E$ ($10^{-5}$ Ry/unit cell)  \\
        \hline
        $10\times 10\times 1$ & -22.9   \\
        $18\times 18\times 1$ & -7.5 \\ 
        $24\times 24\times 1$ & -5.5 \\ 
        $30\times 30\times 1$ & -5.9 \\ 
    \end{tabular}
    \end{center}
    \label{tab:k-points}
\end{table}

\section*{Author contributions (CR{edi}T)}
\textbf{Miranda Naurin:} Conceptualization (equal), Formal analysis (supporting), Investigation (lead), Visualization, Writing - original draft (equal).
\textbf{Sally Aldhaim, Moltas Elliver, Ludwig Hagby, and Didrik Nilsson:} Conceptualization (equal), Investigation (supporting), Writing - review \& editing.
\textbf{Elsebeth Schr\"oder:}  Conceptualization (equal), Data curration, Formal analysis (lead), Funding acquisition, Investigation (supporting), Resources, Supervision, Validation, Writing - original draft (equal).

S.A., M.E., L.H., and D.N.\ contributed equally to this work. 

\section*{Acknowledgment}
Thanks go to D. Orlov (Lund) and P. Maier (Stralsund) for discussions.
The present work is supported by the Swedish Research Council (VR) through Grant No.\ 2020-04997, 
and by Chalmers Area of Advance (AoA) Nano and AoA Materials.
The computations were performed using computational and storage resources at 
Chalmers Centre for Computational Science and Engineering (C3SE), 
and with computer and storage allocations from the 
National Academic Infrastructure for Supercomputing in Sweden (NAISS), under contracts
NAISS2024/3-16, 
NAISS2024/6-432, 
NAISS2025/3-25, and 
NAISS2025/5-484. 

The authors have no conflicts of interest to disclose.

\appendix
\section{Convergence of k-points and energy cut offs\label{param}}

To estimate the k-point sampling needed for a 1$\times$1 surface unit cell, 
 we calculated the total energy of Mg(OH)$_2$ on top of Mg(0001), with all atomic positions locally optimized, 
 and studied the  energy cost $\Delta E$ of lifting the Mg(OH)$_2$ layer 0.1 {\AA} from Mg(0001), 
 with all atoms in fixed positions. 
Results are shown in Table \ref{tab:k-points}.
We find that with $18\times18\times1$ the calculations are sufficiently converged, with 0.02 mRy (0.2 meV) error 
per $1\times1$ surface unit cell, while the $10\times10\times1$ sampling converge the results to 0.15 mRy (2 meV) 
per unit cell. As discussed in the main text, we use the more dense sampling for Mg(OH)$_2$ on Mg(0001) calculations
while for amino acid adsorption the more sparse choice is sufficient. 

With a similar procedure we determined the energy cutoffs $E_\cut$ and $E_\cut^\rho$ for the 
plane wave densities and the charge density,
at $18\times18\times1$ k-point sampling.   
The three sets of values tested were 40/320, 50/400, and 70/560 Ry. 
We concluded that 50/400 Ry is sufficient for the project with the present choice of pseudopotentials and exchange-correlation
approximation.


\end{document}